\documentclass[9pt,twocolumn,twoside]{osajnl}
\usepackage[ justification=justified]{caption} 
\usepackage{multirow}
\usepackage{titlesec}
\graphicspath{{images/}{../images/}}
\usepackage{tabularx}

\journal{optica} 
\setboolean{shortarticle}{false}
\usepackage{lineno}

\newcites{S}{References}

\title{ Continuous-variable quantum passive optical network}

\author[1,$\dagger$, *]{Adnan A.E. Hajomer}
\author[1,2,$\dagger$,**]{Ivan Derkach}
\author[2]{ Radim Filip}
\author[1]{Ulrik L. Andersen}
\author[2]{Vladyslav C. Usenko}
\author[1,***]{Tobias Gehring}

\affil[1]{Center for Macroscopic Quantum States (bigQ), Department of Physics, Technical University of Denmark, 2800 Kongens Lyngby, Denmark}
\affil[2]{Department of Optics, Faculty of Science, Palacky University, 17. listopadu 12, 771 46 Olomouc, Czech Republic}
\affil[$\dagger$]{These authors contributed equally}
\affil[*]{Corresponding authors: * aaeha@dtu.dk, ** ivan.derkach@upol.cz, ***tobias.gehring@fysik.dtu.dk}

\begin{abstract}

Building scalable and secure quantum networks with many users has a high application potential but also holds many practical challenges. A significant stride in this pursuit involves extending quantum key distribution---an information-theoretically secure method for establishing cryptographic keys between two distant users---from a point-to-point protocol implemented on direct optical connections to a quantum access network. Yet, realizations of quantum access networks have, so far, relied on probabilistic or time-sharing strategies. Here, we show theoretically and experimentally that a solution without these constraints can come from the exclusive features of continuous-variable systems. Based on coherent states, we propose continuous-variable quantum passive-optical-network (CV-QPON) protocols, enabling deterministic and simultaneous secret key generation among all network users. We achieve this by leveraging the inherent wave-like property of coherent states split at a beam splitter and electric-field quadrature measurements. We show two protocols with different trust levels assigned to the network users and experimentally demonstrate key generation in a quantum access network with 8 users, each with an 11 km span of access link. Depending on the trust assumptions about users, we reach 1.5 Mbits/s and 2.1 Mbits/s of total network key generation. Demonstrating the potential to expand the network's capacity to accommodate tens of users at a high rate, our CV-QPON protocols offer a pathway toward establishing low-cost, high-rate, and scalable quantum access networks using standard telecom technologies and directly exploiting the existing access network infrastructure. 
\end{abstract}

\setboolean{displaycopyright}{false}

\begin{document}

\maketitle
\section{Introduction}

Quantum key distribution (QKD), the cornerstone of quantum communication, enables two parties to share information-theoretically secure cryptographic keys by exchanging quantum systems over an insecure quantum channel~\cite{pirandola2020advances}. Currently, QKD is advancing towards commercial applications, forming the backbone of quantum networks through point-to-point (PTP) links with trusted nodes \cite{peev2009secoqc, sasaki2011field, wang2014field}. 

Recent advancements have also focused on point-to-multipoint (PTMP) QKD connections, addressing the crucial 'last-mile user access' problem ~\cite{townsend1997quantum, frohlich2013quantum, wang2023experimental}. While PTMP QKD using discrete variable (DV) systems has been proposed for broadcasting channels in passive optical networks (PONs), where a single transmitter is connected to multiple receivers through a passive optical beam splitter, challenges such as the probabilistic nature of user access and the high cost of single photon detectors at each receiver station have limited its applicability~\cite{townsend1997quantum}. As a cost-effective solution, the upstream quantum access network was introduced \cite{frohlich2013quantum}, utilizing a time-multiplexing strategy to share a single photon detector among multiple transmitters. However, this approach significantly limits the secret key rate and becomes increasingly complex with more users due to the time slot allocation~\cite{takeoka2017}. Therefore, it is essential to develop new QKD-based access network protocols to effectively tackle the persistent challenges associated with last-mile user access. 

In this article, we propose continuous variable protocols for quantum passive-optical-networks (CV-QPON) that facilitate deterministic and simultaneous key exchange among all CV-QPON users with information-theoretic security in the presence of Gaussian resources. These protocols extend the scope of CV quantum cryptography from PTP to scalable PTMP networks, a crucial aspect for large-scale deployment. We focus on a downstream CV-QPON topology where a provider (Alice) connects to multiple users (Bobs) via an insecure quantum broadcast channel, potentially under adversary control (Eve). Quantum correlations are established by preparing random coherent states at Alice's station, then simultaneously measured by Bobs. This setup enables independent key generation between Alice and each Bob, thanks to the independent quantum noise experienced by each user and the use of reverse information reconciliation~\cite{grosshans2003quantum}.

Our security analysis encompasses two scenarios: an untrusted protocol, where each Bob views others as potential adversaries, and a trusted protocol, where users collaborate against Eve by relying on a faithful operation of each other. Our trusted protocol uniquely addresses the issue of information leakage due to the residual correlation between users without compromising the secret key's length. This is accomplished by establishing a hierarchical system of trust among users. We demonstrate the feasibility of our proposed protocols through an experimental CV-QPON setup involving eight users, each with an 11 km span of access link.  In both trusted and untrusted scenarios, all users can simultaneously generate independent keys secure against collective attacks in the asymptotic regime, with  $\approx30\%$ improvement in the total network key rate for trusted protocols. Specifically, we achieved total network key rates of 2.1 Mbits/s and 1.5 Mbits/s for trusted and untrusted protocols, respectively. The capacity of our CV-QPON protocol is scalable, allowing it to support more than twice the current number of users, depending on the noise and channel transmittance. Additionally, CV-QPON offers a cost-effective solution as it utilizes standard telecommunications technology, enabling it to be effortlessly integrated into existing access networks.   

\section{Network architecture and operation}

\begin{figure*}
    \centering
    \includegraphics[width=0.99\linewidth]{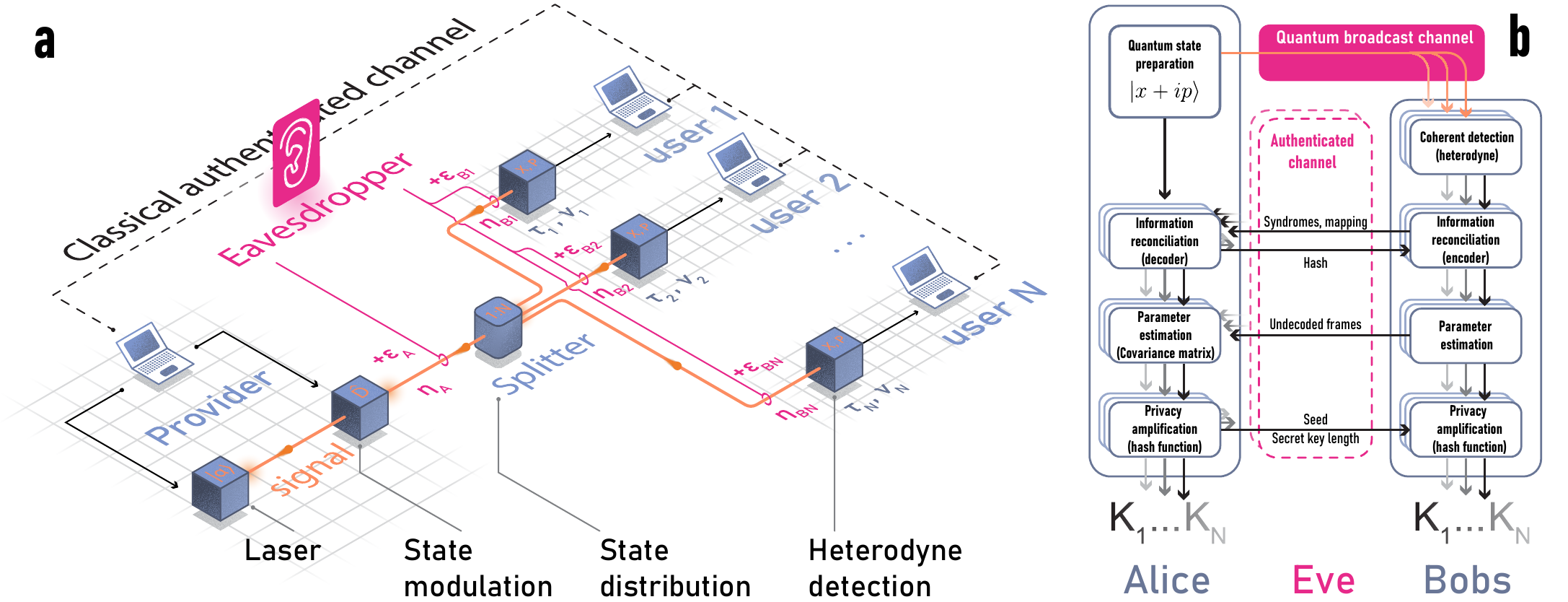}
    \caption{\textbf{Continuous-variable quantum passive optical network (CV-QPON) architecture and operation.}~(a) The provider encodes key information onto quadratures of coherent signal state. This state is broadcast to $N$ users through an insecure quantum channel, under Eve's potential control. The channel's properties include transmittance $\eta$ and Eve-induced excess noise $\epsilon$. Users employ coherent detection, decoding key information from the coherent state's quadrature values ($x$ and $p$). The detectors are characterized by the quantum efficiency $\tau$ and the electronic noise $v$. All classical communications occur via a classical authenticated channel in a centralized manner. \textbf{(b)} The protocol encompasses a quantum prepare-and-measure phase, followed by data processing. After these steps, Alice and each user will share a symmetric key K$_l$, which can be used for cryptographic tasks. }
    \label{fig:illustration}
\end{figure*}

Figure~\ref{fig:illustration}(a) shows the network architecture of CV-QPON, which is a standard telecom access network topology favored for its high capacity and energy efficiency~\cite{senior2009optical}. Within this network architecture, the nodes are classified based on their distinct roles and functionalities:
\begin{itemize}
\item \textit{Provider} (Alice): Generates and randomly modulates quantum states to establish quantum correlations for secret key generation.
\item \textit{User} (Bob): Performs heterodyne detection on the received optical mode.
\item \textit{Splitter} ($1:N$): A passive component forms a quantum broadcasting channel that connects $N$ users to the QPON infrastructure and evenly distributes quantum correlations among them. 
\end{itemize}

In addition, authenticated classical channels are established between the provider and each user. This setup ensures that all classical communication is centralized, i.e., the users cannot communicate among themselves. In the following we will use both terms 'Bob (B)' and 'the user' synonymously.
\par The key generation process in CV-QPON consists of a series of rounds $k\in(1,M)$, each comprised of the following steps:  \\
\textbf{Preparation:} Alice draws two random variables $x(p)_k$ from independent zero-mean Gaussian distributions $\mathcal{N}(0,V_{x(p)})$ to encode information into a coherent signal state by means of the modulation process. The quadrature variance of the overall generated state is $1+V_{x(p)}$. The preparation station is assumed to be trusted, meaning that it neither leaks information to an eavesdropper nor allows an eavesdropper to control noise within the station.\\
\textbf{Distribution:}  The quantum states are transmitted through an untrusted quantum channel, fully controlled by an eavesdropper with transmittance $\eta_A$, to a splitter, where the laser beam is divided and sent to each user through individual untrusted quantum channels with transmittances $\eta_{B_l}$. The channel is modeled by passive linear optical elements, and 
the total transmittance is given by $\eta_l=\eta_A\eta_{B_l}/N$. Here, we distinguish between total transmittance  $\eta_l$, which covers the entire link, and the segment-specific channel transmittance $\eta_A\eta_B$.  Each link subjects the quantum states to varying levels of noise, with the total excess noise received by each user given by  $\varepsilon_{l}=\varepsilon_{A}\eta_{B_l}/N+\varepsilon_{B_l}$. \\
\textbf{Detection:}  Each user measures the incoming quantum states, monitoring the level of electronic noise (with quadrature variance $\nu_l$) and detection efficiency $\tau_l$.\\ 
\textbf{Post-processing:} After $M$ rounds, Alice engages in data processing with each user over authenticated classical channels, as depicted in Fig. \ref{fig:illustration}(b). This includes information reconciliation, parameter estimation, and privacy amplification. Unlike PTP settings, where each round $k$ is dedicated to a single user $l$,  in CV-QPON protocols, Alice processes the data in parallel by replicating sequences $x(p)_{1,\cdots M}$ to generate $N$ independent secret keys.\\

In the CV-QPON, it is also possible to split the quantum states into $N$ unequal parts. This allows prioritization of certain users or meeting demands of larger keys for preferred services. Nonetheless, the primary focus of this work is on maximizing the number of users that can be supported simultaneously, adhering to the principles of net neutrality. The following section will delve into the protocols that can be implemented within CV-QPON, particularly those facilitating simultaneous key establishment between Alice and each Bob.

    \section{CV-QPON protocols}
The asymptotic key rate is deemed to be secure if the lower bound on the difference between the mutual information of trusted parties $I_{AB_l}$, and accessible information of Eve on measurement of the reference side $\chi_{EB_l}$ remains positive~\cite{devetakwinters2005}:
    \begin{equation} \label{eq:key}
        K_l(\eta,\varepsilon)=\text{max}\left[0,\beta_l I_{AB_l} - \chi_{EB_l}\right],
    \end{equation}
where $\eta$ and $\varepsilon$ are channel parameters, $\beta_l$ is the efficiency of information reconciliation. Both mutual information $I_{AB_l}$ and Holevo bound $\chi_{EB_l}$ are determined by the covariance matrix of the overall shared multipartite state. For the sake of simplicity of notation, we omit detection efficiency and electronic noise. However, they are incorporated in the respective covariance matrices and security analysis. For further details see supplementary materials.\par
    \subsection{Time-sharing approach}
 The most basic method to manage network access among users is known as time-sharing~\cite{frohlich2013quantum}. In this approach, each round  $k$ is allocated to a specific user $l$.  However, this time-sharing PTP QKD protocol faces a significant limitation in key rate as the number of users in the network increases. This is because only a fraction of the rounds, specifically  $M/N$ rounds, are designated for a key generation for each user.  Under the assumption that  all links between the splitter and users have the same losses $\eta_{B_l}=\eta_{B}$ and noise $\varepsilon_{B_l}=\varepsilon_{B}$, the \textit{total secret key rate} generated within the network can be expressed as, 
    \begin{equation}
       K^{TS}_\Sigma=K(\eta_l,\varepsilon_{l}),
    \label{eq:timesharing}
    \end{equation}
which is equal to the standard PTP key rate with a single user over a channel with parameters $\eta_l,\varepsilon_l$ \cite{weedbrook2004quantum}. The time-sharing protocol is particularly suitable for DV QKD-based access networks, where the key is generated by single-photon signal states, and all users time-share the single-photon detector~\cite{frohlich2013quantum}. However, CV coherent states, whose amplitudes can be split into different modes,  enable {\it simultaneous} and \textbf{ \it deterministic} key distribution with different users, and we use these advantages in the following protocols.

    \subsection{Untrusted broadcast protocol} \label{sec:untrusted-protocol}
Due to the multi-photon nature of the coherent state and the use of coherent detection in CV-QPON, detection events will occur for all users in each round of the protocol. Despite the broadcasting of the same coherent state across the CV-QPON, each user, after $M$ rounds, obtains measurement outcomes that are unique, yet weakly correlated. This uniqueness arises from independent quantum noises affecting each user differently.  Through the application of reverse reconciliation~\cite{grosshans2003quantum}, Alice can concurrently generate $N$ keys using the measurement result of each user as a reference. After undergoing the privacy amplification process, these keys become completely independent.  It is critical to ensure that the cost of privacy amplification is sufficient to decouple the final key $K_l$ from Eve \textit{and} all other users. \par
To assure the key independence within the network, one can assume that the fraction, $(N-1)/N$,  of the split signal is intercepted by Eve, instead of being distributed to $(N-1)$ users. This necessitates each user to operate under the presumption that other users may collaborate with Eve.  By adopting this assumption, an upper bound can be established on Eve's information.  Consequently, under this framework, the total network key rate can be quantitatively defined as:
    \begin{equation}
        K^{U}_\Sigma=\sum^{N}_{l=1} K_l = N\times K(\eta_l,\varepsilon_{l}).
    \label{eq:untrusted-key}
    \end{equation}
This approach invariably offers an advantage over the time-sharing protocol as all $M$ rounds are designated for key generation. The concept of this untrusted protocol was theoretically explored in~\cite{huang2021realizing}. However, this study made a specific assumption about the scaling down of channel-related excess noise with an increase in the number of users, thereby overestimating the network's capacity. Instead, in this work, we take advantage of the multi-user nature of the broadcasting protocol and benefit from the dependable operation of network users.

    \begin{figure*}[t]
        \centering
        \includegraphics[width=.9\linewidth]{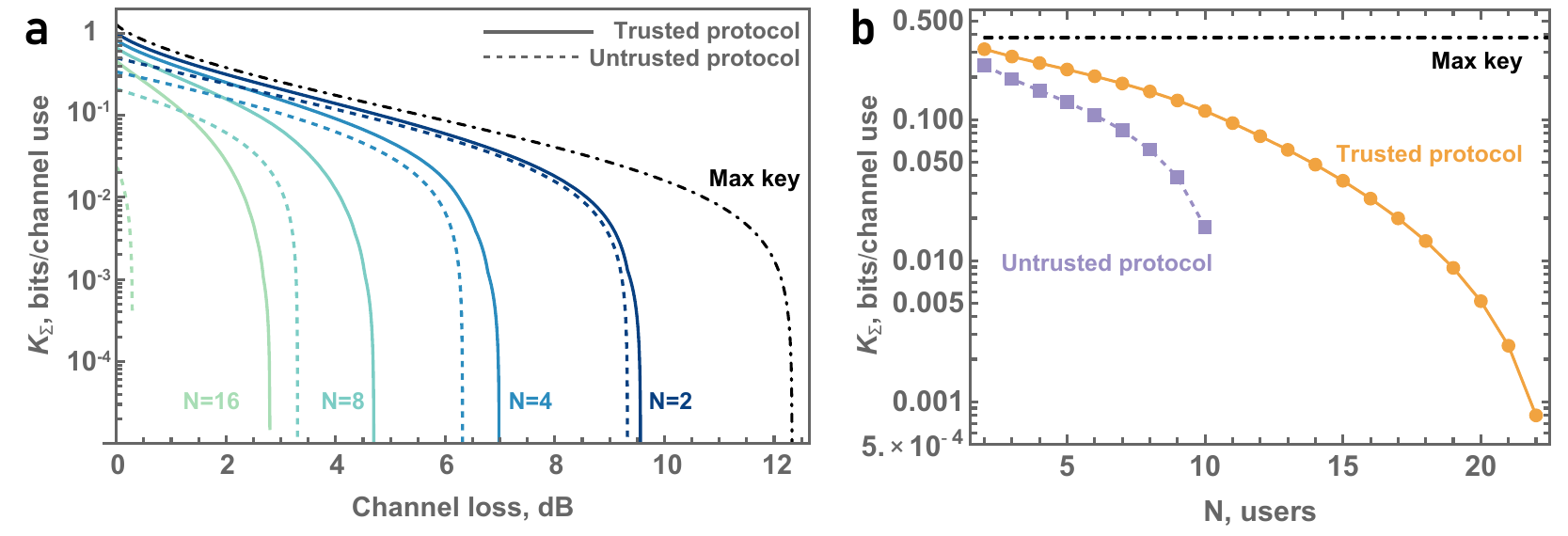}
        \caption{\textbf{Performance and capacity analysis of CV-QPON protocols.} Comparison of the theoretical performance of untrusted protocol (dashed), trusted protocol (solid) and maximal achievable key (dot-dashed) in terms of the lower bound on the total network key rate $K_\Sigma$ dependency on \textbf{(a)} the channel loss $\eta_A\eta_B$ with different number of users $N=2,\,4,\,8,\,16$; \textbf{(b)} number of connected users at fixed channel loss $\eta_A\eta_B= -2$ dB. Parameters: reconciliation efficiency $\beta=95\%$ \cite{zhang2020long, mani2021multiedge}, modulation variance $V_{x(p)}= 4$ SNU (which is optimal for a large number of users), detector efficiency $\tau=86$\%, electronic noise $\nu=2\%$ shot-noise unit (SNU), excess noise at channel output $\varepsilon=0.5\%$ SNU.}
    \label{fig:2}
    \end{figure*}
    
    \subsection{Improving network performance} \label{sec:naive-protocol}
In PTP CV-QKD protocols, the security level can be defined based on the degree of trust assigned to different parts of the system, specifically, those parts that can potentially be under/beyond Eve's control~\cite{scarani2009security, jouguet2012analysis}. Typically, a higher security level implies fewer assumptions about Eve's ability to access and control the system. This, in turn, influences both the achievable key rate and the secure distance that can be reached. However, some deviations from nominal performance, e.g., imperfect detection including non-unity quantum efficiency and electronic noise, can be regarded as trusted, provided that the respective equipment is thoroughly characterized and monitored. These deviations then do not enhance Eve's knowledge about the key. \par
In this work, we extend this notion of trust among QPON users. Specifically, when user $B_i$ trusts user $B_j$, $B_i$ assumes that $B_j$ successfully receives and measures the $1/N$ portion of the signal, instead of it being intercepted by Eve. This shift in perspective enables $B_i$ to attribute the corresponding signal loss to an overall trusted multipartite state, rather than to Eve's intervention. Consequently, this lowers the accessible information of Eve $\chi_{EB_i}$ while maintaining the mutual information between the provider and $B_i$, denoted as $I_{AB_i}$. Thus, it enhances the overall key rate.

To obtain $\chi_{EB_i}$, it is necessary to  reconstruct a covariance matrix corresponding to the trusted state, which contains modes of Alice, $\text{Bob}_i$ and $\text{Bob}_j$, along with the  purifications of realistic detectors 
 \cite{usenko2016trusted}. The reconstruction of this matrix involves estimating parameters for both users  ($\eta_i$, $\varepsilon_i$, $\tau_i$, $\nu_i$ and $\eta_j$, $\varepsilon_j$, $\tau_j$, $\nu_j$), a task performed by Alice. For a detailed description of the modeling of this trusted system, please refer to the supplementary material. \par
On the other hand, misplaced assumptions can undermine the security of the entire network. Suppose all users have full trust in each other's faithful operation. In an attempt to establish keys, Alice reconstructs a full covariance matrix with $N$ users. She presumes that Eve can only access information from ancillary modes before and after the splitter with total number of modes equal to $N+1$ . However, during information reconciliation, each user transmits a syndrome related to their measured data, as shown in Fig.~\ref{fig:illustration}(b). This allows Alice to reconcile her data string based on the reference user's measurement. Since all users are correlated, every syndrome provides non-negligible information about non-reference user's measurements as well. This issue is further amplified by the inefficiency of reconciliation algorithms $\beta\in[0,1)$, necessitating sending larger syndrome than the theoretically required minimum. Hence, if all users simultaneously attempt to minimize the cost of privacy amplification, they might significantly underestimate Eve's information,  endangering the network's security.  

One way to solve this issue is to use part of the generated key from the previous QKD session to encrypt the syndrome with a one-time pad \cite{bian2023high}. However, the exact encryption cost in terms of reserved key volume that would be sufficient to preserve the security must be determined in advance. Additionally, the amount of pre-shared key needed to initiate the protocol also increases significantly. Furthermore, the irreversible property of the protocol, i.e., each QKD session is independent of the others, no longer holds in this context. \par

    \subsection{Trusted broadcast protocol} \label{sec:trusted-protocol}
We introduce a new protocol that outperforms untrusted protocols in terms of network key rate while avoiding disclosing information regarding other network users through the syndromes.  Upon completing $M$ rounds of the protocol, Alice initiates key distillation with multiple users simultaneously. Starting with $B_1$, who considers $B_2\cdots B_N$ as \textit{untrusted} party, effectively under Eve's control, he opts for the maximum privacy amplification penalty. However, this also implies that no other Bob can threaten the security of the final key $K_1$. Knowing this, $B_2$ can now classify $B_1$ as a trusted user,  as there is no threat to the security of $K_1$ from his actions, though he still regards $B_3\cdots B_N$ as untrusted users.  This strategy enhances the secure key rate, as in a simplified scenario with identical parameters in all $N$ channels, $K_1<K_2$. Following this pattern, each successive user trusts all preceding users, accruing an additional key advantage progressively. By varying the order of trust among users in each session,  we can optimize the network key rate gain $K^T_\Sigma\geq K^U_\Sigma$, where the equality holds only when no key can be established, without violating the security of individual users. 

In scenarios where certain users are unable to generate keys, it does not necessarily indicate a comprehensive compromise of the network's ability to generate secret key with those users. Indeed they can still attain a positive key rate by adopting a more substantial degree of trust. Figure~\ref{fig:2} delineates this principle, showing that under conditions of higher loss or an increased number of users, the untrusted broadcast protocol fails to maintain a non-zero key rate. In contrast, the trusted protocol continues to yield a positive key rate for some users even under these challenging conditions. Furthermore, Fig.~\ref{fig:2} provides a comparative analysis of the performance of two broadcast protocols against the maximal key rate achievable through a PTP protocol. This analysis is conducted under identical conditions, of channel loss, $\eta_A\eta_B$, and equivalent levels of excess noise, $\varepsilon$, at the output of the quantum channel. The comparison indicates the possibility of an optimal CV-QPON protocol capable of further improving the total network key or even saturating the PTP key rate. Notably, the larger the network the bigger the quantitative improvement of the key rate when users are assumed to be trusted.\\

\section{Experimental implementation}

    \begin{figure*}[t]
    \centering
        \includegraphics[width=.8\linewidth]{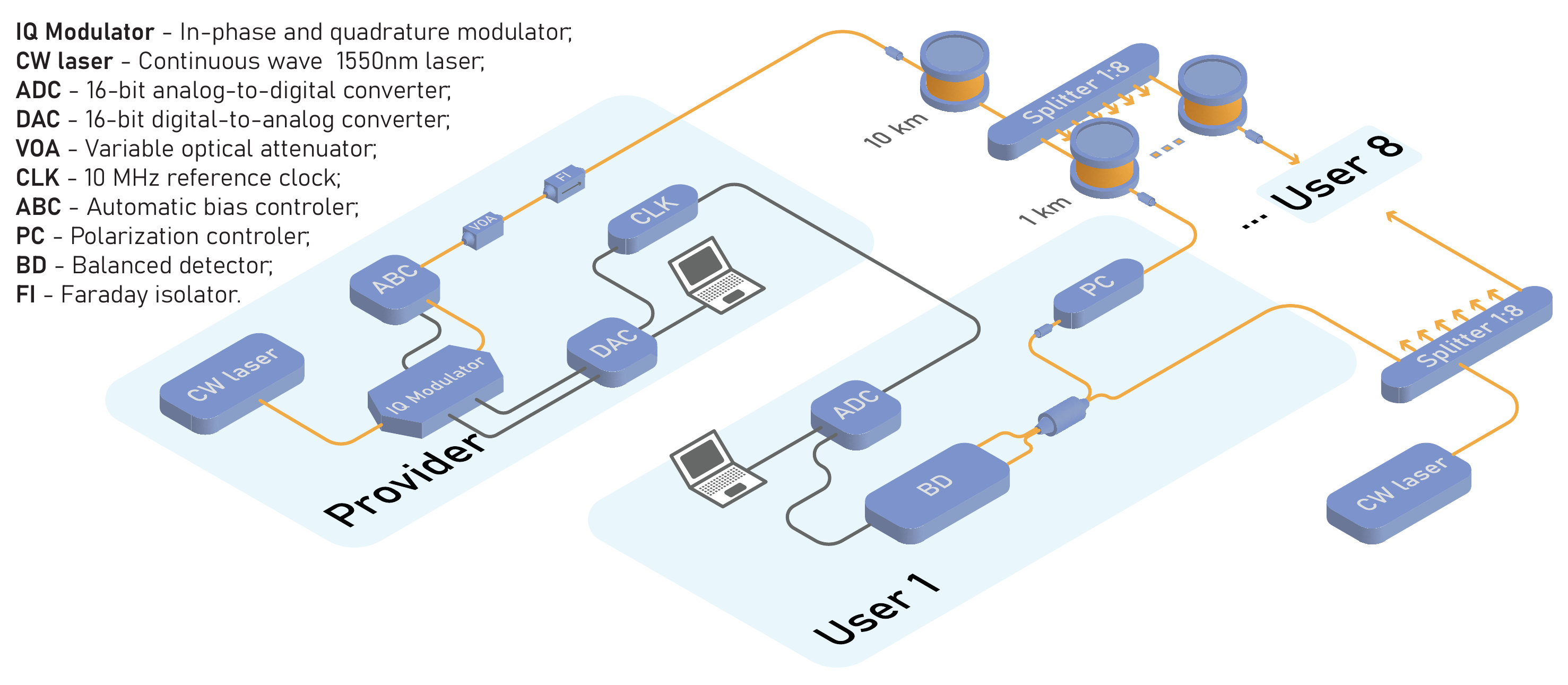}
        \caption{\textbf{Experimental setup of CV-QPON}.  The setup involves a provider, Alice, generating a coherent state in the single sideband of the optical carrier, utilizing a CW laser and an IQ modulator driven by a DAC and automatic ABC. The system connects Alice to eight users via a passive optical splitter and fiber spools. Each user employs RF heterodyne detection, utilizing an independent CW laser as an LO shared among users, a BD, and a polarization controller PC to adjust the quantum signal's polarization. The detected signals are then digitized using 1 GSample/s DAC card, synchronized to the DAC with a CLK.}
    \label{fig:experimental-scheme}
    \end{figure*}

    \begin{figure*}
        \centering
        \includegraphics[width=.7\linewidth]{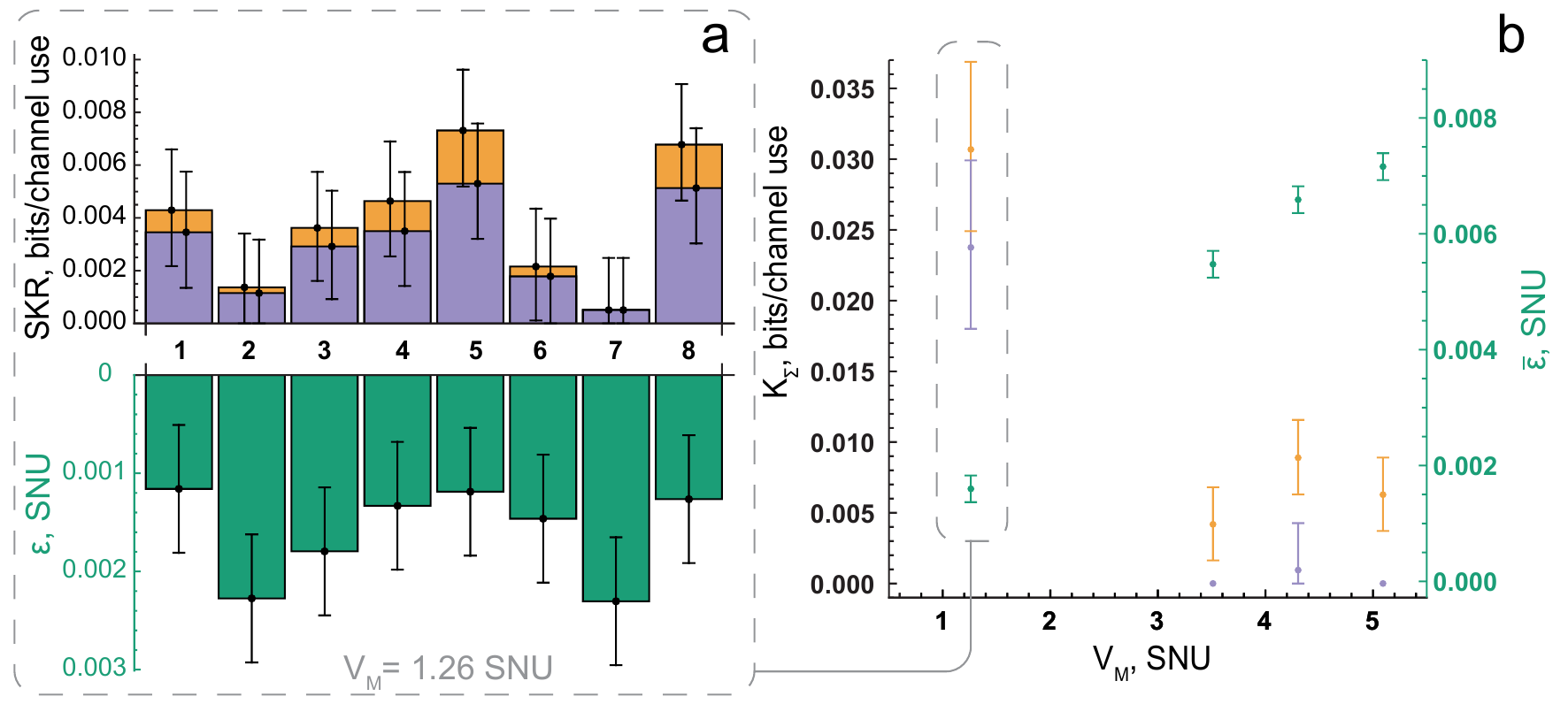}
        \caption{\textbf{Performance of CV-QPON protocols - experimental results}. (a) The top section illustrates the secure key rate (SKR) achieved by each user with modulation variance of $V_M = 1.26$ SNU  for the untrusted protocol (violet bars) and trusted protocol (orange bars), with assumed reconciliation efficiency of  $\beta=95\%$. The bottom section displays the corresponding excess noise measured at the user stations. Noise error bars are determined by Gaussian confidence intervals with 6.5 standard deviations, and failure probability of $\delta=10^{-10}$ \cite{ruppert2014long}. SKR error bars are given by respective lowest/highest noises and transmittances within confidence intervals. (b) Total network key rate $K_\Sigma$ and mean excess noise $\bar{\varepsilon}$ for all eight users at different modulation variance settings.}
    \label{fig:experimental-results}
    \end{figure*}

We experimentally demonstrated the CV-QPON. Figure~\ref{fig:experimental-scheme} shows the schematic of the experimental setup. This setup consists of one transmitter connected to eight receivers through an 11 km quantum broadcast channel, made of a 1:8 passive optical beam splitter and single mode fibers (SMF). The provider, Alice, produces an ensemble of coherent states. This process involves two main components: a digital signal processing (DSP) module and an optical module. In the DSP module, the complex amplitude of each coherent state was formed by drawing random numbers from  Gaussian distributions, obtained from a vacuum-based quantum random number generator (QRNG)~\cite{Gehring2021qrng}. The quantum symbols, drawn at a rate of 100 MBaud, were upsampled to 1 GSample/s. Subsequently,  they were pulse-shaped using a root-raised cosine filter with a roll-off factor of 0.2. The resulting baseband signal was frequency-shifted to center around 170 MHz, aiding in single-sideband modulation. Additionally, a 270 MHz pilot tone was frequency-multiplexed with the passband signal to facilitate carrier phase recovery. The corresponding electrical signal was generated using a digital-to-analog converter (DAC) operating at a sampling rate of 1 GSample/s. 

In the optical module, a 1550 nm continuous wave (CW) laser with a linewidth of 100 Hz was used as an optical source. This laser was modulated by an in-phase and quadrature (IQ) modulator, driven by the DAC.  The IQ modulation was set to operate in optical single-sideband carrier suppression mode. To achieve this, an automatic bias controller (ABC) was used to control the direct current bias voltages applied to the IQ modulator. Following the IQ modulator, a variable optical attenuator (VOA) was used to finely adjust the modulation variance of the generated thermal state. The quantum signal was then sent to eight receivers through the quantum broadcast channel. Each receiver, on average, experienced a physical loss of  approximately $-13.8$ dB.    

At the receiver ends, each user used coherent detection to measure the incoming coherent states. This involved implementing radio frequency (RF) heterodyne detection, which mixes the quantum signal with a local oscillator (LO) signal in a balanced beamsplitter. The LO was generated by an independent, free-running CW laser, which had a frequency offset of $\approx 300 $ MHz relative to Alice's laser. Due to a limitation of available equipment, all eight receivers shared the same LO, split by another 1:8 beam splitter. A manual polarization controller (PC) was then used to align the polarization of the quantum signal with that of the LO to maximize the visibility of interference fringes. The interference pattern was detected and digitized using a broadband balanced detector (BD) with a bandwidth of $\approx250$ MHz and a 1 GSample/s analog-to-digital converter (ADC). To emulate the actual scenario of the network setting,  each pair of receivers was associated with an independent workstation, each of which was equipped with two-channel ADC cards. These ADCs were clock synchronized with the DAC using a 10 MHz reference clock. The workstations were connected through a local area network, and the entire setup was controlled through a Python-based framework, enabling autonomous modulation and data acquisition.   

The users performed three types of measurements: the quantum signal, the vacuum noise (transmitter's laser off, LO laser on), and the electronic noise (transmitter's laser off, LO laser off). These measurements were carried out consecutively, and divided into frames, each containing $10^7$ ADC samples. For the modulation variance calibration, a back-to-back measurement was conducted by directly connecting one of the receivers to Alice's station using a short fiber patchcord. The clearance on the quantum band of each user was, on average, $\approx15~$dB. Following these measurements, the receivers started the process of recovering their quantum symbols using an offline DSP module.

The DSP technique for quantum symbols recovery involves several steps~\cite{hajomer2024long}. First, it applies a whitening filter to remove any correlation between the quantum symbols caused by the detector's imperfect transfer function. Next, it utilizes a pilot-aided carrier phase recovery,  enhanced by employing a machine-learning method based on an unscented Kalman filter~\cite{chin2021machine}. This is followed by achieving temporal synchronization through cross-correlating with predefined reference symbols. The final stages included matched filtering and downsampling to the symbol rate.

Upon completion of the prepare-and-measure phase, users progressed to the subsequent stages of the CV-QPON protocols. The initial step is information reconciliation, a critical process wherein users adopted a multi-dimensional (MD) reconciliation  approach~\cite{leverrier2008multidimensional}. This method relied on multi-edge-type low-density-parity-check (MET-LDPC) error-correcting codes with a rate of 0.01~\cite{mani2021multiedge}. To enhance reconciliation efficiency, rate-adaptive techniques were integrated~\cite{wang2017efficient}. For more detailed information, readers are directed to the supplementary materials. \\ Subsequently, Alice undertook the task of parameter estimation. Within the framework of an untrusted protocol, Alice constructed covariance matrices for each user to compute the key rate. Conversely, for the trusted broadcast protocol, a covariance matrix describing the overall shared state was reconstructed. Then the users' trust sequence was assigned in the ascending order based on their key rates obtained from the untrusted broadcast protocol (\ref{sec:untrusted-protocol}). Specifically, the user with the lowest key rate perceived all others as untrusted, whereas the user with the highest key rate considered all others as trusted, thereby elevating the key rate even more. Such a strategy has been determined to maximize the network key gain. Interestingly, inverting this order enables the user with the lowest key rate to have the most significant enhancement. The final step in both protocols involved implementing privacy amplification~\cite{Tang2019}.  


\section{Results}
In our network architecture, we implemented a local LO (LLO) scheme to rule out side-channel vulnerabilities and simplify the network setup. A main challenge encountered with this approach is the laser phase noise, mainly arising from the use of independent lasers at the provider and user stations. Although all users share the same LO, we assume that the excess noise affecting each user is independent. This is because each user has a separate fiber channel, and each channel introduces its own independent phase noise. However, in general, noise correlations could influence key rate performance depending on other network parameters. \par

Optimizing the modulation variance, $V_M$, for a specific reconciliation efficiency, $\beta$, can theoretically enhance protocol performance. However, in practice, increasing  $V_M$ leads to higher excess noise, attributable to the residual phase noise. This correlation is evident from the linear scaling of the excess noise with $V_M$, as highlighted in Fig.~\ref{fig:experimental-results}(b). Consequently, this complicates the determination of an optimal range for $V_M$ \cite{hajomer2024long} for a given $\beta$. Thus, to facilitate concurrent network access, we chose a $V_M$ of 1.26  shot-noise unit (SNU) to align with both the MET-LDPC code rate and the levels of expected excess noise. \par

Figure~\ref{fig:experimental-results} compares the total key rates of both trusted and untrusted  CV-QPON protocols at various modulation variances, respective measured mean excess noise $\bar{\varepsilon}$,  with a $\beta$ of 95\% \cite{zhang2020long,mani2021multiedge}. It shows that minor alterations in channel parameters can significantly impact key rates, underlining the importance of carefully managing modulation variance to support simultaneous key generation for all users. Compared to the untrusted protocol, our developed trusted protocol (indicated by orange bars)  increases the total network key rate by  $\approx30\%$ and enables a positive key rate at higher $V_M$, where phase noise is more pronounced.


Table~\ref{tab:results} summarizes the experimental parameters for key generation, covering the full protocol implementation, including information reconciliation and privacy amplification. Alice generated $M=10^8$ coherent states with a modulation variance of $V_M=$1.26 SNU. The trusted loss at user stations was $\tau=0.685$.  Information reconciliation yielded various efficiencies $\beta$ and frame error rate (FER),  reflecting the different received signal-to-noise ratios (SNR)  for each user, due to unique channel transmittances $\eta_l$. 

In the untrusted CV-QPON protocol scenario, Bob$_5$ achieved a maximum key rate ($K^U$) of 375.4 kbit/s, thanks to low excess noise level and high channel transmittance. Notably, under the trusted protocol, the key rate for the same user was increased by 46\%,  underscoring our CV-QPON protocols' capacity to support simultaneous key generation and enhance network performance. Adopting finite-size effect analysis of PTP protocols \cite{leverrier2010finite, ruppert2014long}, we can establish secure keys with four users (no. 1, 4, 5 and 8) using the trusted protocol by assuming reconciliation $\beta=96\%$ and protocol failure probability $\delta=10^{-10}$. However, it is important to note that broadcasting protocols require a more comprehensive channel estimation and finite-size effects analysis from a multi-user perspective.

\begin{table}[t]
\caption{\textbf{Summary of experimental parameters and CV-QPON protocols performance}. $\eta_l$: channel transmittance, $V_l$: electronics noise, $\varepsilon:$ excess noise (at the channel output), $\beta$: information reconciliation efficiency, FER: frame error rate, $K^{U(TS)}:$ untrusted(trusted) key rate.}
    \centering
    \resizebox{\hsize}{!}{
    \begin{tabular}{|l|l|l|l|l|l|l|}
\hline
User          &$\eta_l$    &$V_{l}$, mSNU & $\varepsilon$, mSNU & $\beta$, \%& FER,\%& $K^{U(TS)}$~kbits/s \\ \hline
$\text{Bob}_1$& 0.0369& 51.24& 0.794& 90.79 & 4.5 & 242.8~(322.6)\\ \hline
$\text{Bob}_2$& 0.0424& 52.76& 1.558& 93.23 & 43  & 40.4~(53.1)\\ \hline
$\text{Bob}_3$& 0.0439& 55.42& 1.23 & 91.37 & 22.3 & 154.3~(208.9)\\ \hline
$\text{Bob}_4$& 0.0397& 49.74& 0.912& 91.5  & 15.3 & 227.1~(323.5)\\ \hline
$\text{Bob}_5$& 0.0461& 60.14& 0.814& 91.44 & 13.6 & 375.4~(549.2)\\ \hline
$\text{Bob}_6$& 0.0337& 53.14& 1.002& 91.9 & 21.5 & 92.01~(121.2)\\ \hline
$\text{Bob}_7$& 0.0398& 75.18& 1.578& 94.8  & 55.4 & 20.73~~(20.73)\\ \hline
$\text{Bob}_8$& 0.0463& 52.66& 0.866& 90.78 & 9.5 & 360.5~(509.6)\\ \hline
     
    \end{tabular}
}
\label{tab:results}
\end{table}

\section{DISCUSSION}
In CV quantum cryptography, the quantum information is encoded in the quadratures of the electromagnetic field of light and subsequently decoded via coherent detection. This offers the advantage of employing cost-effective, standard telecom components operating at room temperature and a high rate over metropolitan distances compared to discrete variable protocols~\cite{hajomer2023continuous}. Despite these advantages, the application of CV quantum cryptography has been predominantly confined to point-to-point connections and niche applications in dedicated high-security networks. In our work, we have extended the scope of CV quantum cryptography beyond the point-to-point paradigm to encompass quantum access networks. This expansion has been achieved by introducing continuous-variable quantum passive-optical-network protocols. 

The developed CV-QPON protocols are designed to enable simultaneous key generation— a distinct feature of CV-QPON— and ensure compatibility with conventional downstream access network architectures.  Depending on the assigned trust level to network users, we have defined both untrusted and trusted CV-QPON protocols, and have experimentally validated the feasibility of these protocols within a CV-QPON setup based on a LLO scheme. Our network facilitates concurrent access to eight users over an 11 km access link, with the potential to scale up the number of users based on the excess noise and the channel loss. We have shown that the trusted protocol significantly enhances the overall network key rate performance and that it is particularly advantageous in scenarios where each user experiences different levels of channel loss. 

Unlike the successive quantum state merging protocols \cite{takeoka2017}, our trusted protocol is compatible with an arbitrary selection of Gaussian states, i.e., both coherent and squeezed states, and offers the flexibility to incorporate effects of trusted detectors and various side channels \cite{derkach2016,jain2021modulation}. A unique feature of this protocol is its strategy for removing information leakage stemming from residual correlations among users and the dissemination of information reconciliation syndromes. This is achieved by establishing a lower bound on the key rate through a hierarchical trust model, obviating the necessity for additional syndrome encryption~\cite{bian2023high} while maintaining the secrecy and irreversibility of the protocol's operations. Compared to time-sharing approaches ~\cite{wang2023experimental, frohlich2013quantum}, our CV-QPON protocols demonstrates a definitive advantage in terms of key rate and the capability for concurrent key generation. Nonetheless, there remains scope for further enhancements. Theoretically, identifying a CV-QPON protocol that can saturate the channel's capacity is imperative for achieving optimal performance. Addressing the finite-size effects in the current trusted protocol and extending the security proof to encompass discrete modulation are pivotal for enhancing practical implementation and achieving higher rates through the use of high-speed components~\cite{hajomer2023continuous}. Furthermore, reducing excess noise through improved laser technologies and the development of MET-LDPC codes optimized for optimal modulation variance are essential for expanding network capacity and enhancing total network key rates.

In conclusion, our CV-QPON protocols offer a cost-effective, practical solution that seamlessly integrates with standard telecom network infrastructures, thereby facilitating the progression toward a comprehensively interconnected quantum network, such as the European quantum communication infrastructure (EuroQCI).

\begin{backmatter}
\vspace{0.5cm}
\bmsection{Data availability} All data needed to evaluate the conclusions in this paper are present in the paper and/or the Supplementary Materials.

\smallskip

\bmsection{Acknowledgments} We thank Masahiro Takeoka, Akash nag Oruganti, Florian Kanitschar and Christoph Pacher for fruitful discussions and TDC NET A/S for the equipment loan. This project was funded within the QuantERA II Programme (project CVSTAR) that has received funding from the European Union’s Horizon 2020 research and innovation program under Grant Agreement No 731473 and 101017733; from the European Union’s Digital Europe programme under Grant agreement No 101091659 (QCI.DK); from the European Union’s Horizon Europe research and innovation programme under the project ``Quantum Security Networks Partnership'' (QSNP, grant agreement no. 101114043). AH, ULA and TG acknowledge support from Innovation Fund Denmark (CryptQ, 0175-00018A) and the Danish National Research Foundation, Center for Macroscopic Quantum States (bigQ, DNRF142).  AH and TG acknowledge funding from the Carlsberg Foundation, project CF21-0466. ID acknowledges support from the project 22-28254O of the Czech Science Foundation. RF acknowledges project 21-13265X of the Czech science Foundation. RF and VCU acknowledge the project CZ.02.01.01/00/22{\_}008/0004649 (QUEENTEC) of the Czech MEYS.  VCU acknowledges the project 21-44815L of the Czech Science Foundation. 

\smallskip

\bmsection{Competing interests} The authors declare no competing interests.

\smallskip

\bmsection{Author contributions statement} A.A.E.H.\ and I.D.\ contributed equally as first authors. A.A.E.H.\ designed the experiment, implemented the DSP routine, and performed the overall data processing and analysis under the supervision of T.G.  I.D.\ developed the overall theoretical model and security analysis under the supervision of V.C.U. and feedback from R.F. A.A.E.H.\ and I.D.\  wrote the manuscript with input from T.G.\ and V.C.U.\ A.A.E.H. and T.G. conceived the experiment. R.F.,\ U.L.A.,\ V.C.U.\, and T.G. supervised the project in different parts. All authors were involved in discussions and interpretations of the results.
\smallskip

\end{backmatter}

\bigskip

\bibliography{lib}

\bibliographyfullrefs{lib}

\ifthenelse{\equal{\journalref}{aop}}{%
\section*{Author Biographies}
\begingroup
\setlength\intextsep{0pt}
\begin{minipage}[t][6.3cm][t]{1.0\textwidth} 
  \begin{wrapfigure}{L}{0.25\textwidth}
    \includegraphics[width=0.25\textwidth]{john_smith.eps}
  \end{wrapfigure}
  \noindent
  {\bfseries John Smith} received his BSc (Mathematics) in 2000 from The University of Maryland. His research interests include lasers and optics.
\end{minipage}
\begin{minipage}{1.0\textwidth}
  \begin{wrapfigure}{L}{0.25\textwidth}
    \includegraphics[width=0.25\textwidth]{alice_smith.eps}
  \end{wrapfigure}
  \noindent
  {\bfseries Alice Smith} also received her BSc (Mathematics) in 2000 from The University of Maryland. Her research interests also include lasers and optics.
\end{minipage}
\endgroup
}{}

\pagebreak
\onecolumn
\begin{center}
\textbf{\huge Supplementary Materials for\\ \vspace{0.5cm} Continuous-variable quantum passive optical network}\\ \vspace{0.5cm} Adnan A.E. Hajomer$^*$$^\dagger$, Ivan Derkach$^{**}$$^\dagger$, Radim Filip, Ulrik L. Andersen, Vladyslav C. Usenko, Tobias Gehring$^{***}$ \\ \vspace{0.5 cm}
$\dagger$ These authors contributed equally\\
Corresponding authors: * aaeha@dtu.dk, ** ivan.derkach@upol.cz, *** tobias.gehring@fysik.dtu.dk
\end{center}
\setcounter{equation}{0}
\setcounter{section}{0}
\setcounter{figure}{0}
\setcounter{table}{0}
\setcounter{page}{1}

\makeatletter
\renewcommand{\theequation}{S\arabic{equation}}
\renewcommand{\thefigure}{S\arabic{figure}}
\renewcommand{\thetable}{S\arabic{table}}
\renewcommand{\bibnumfmt}[1]{[S#1]}
\renewcommand{\citenumfont}[1]{S#1}


    \section{Constructing the covariance matrix}  
   
\begin{figure}
    \centering
    \includegraphics[width=0.7\linewidth]{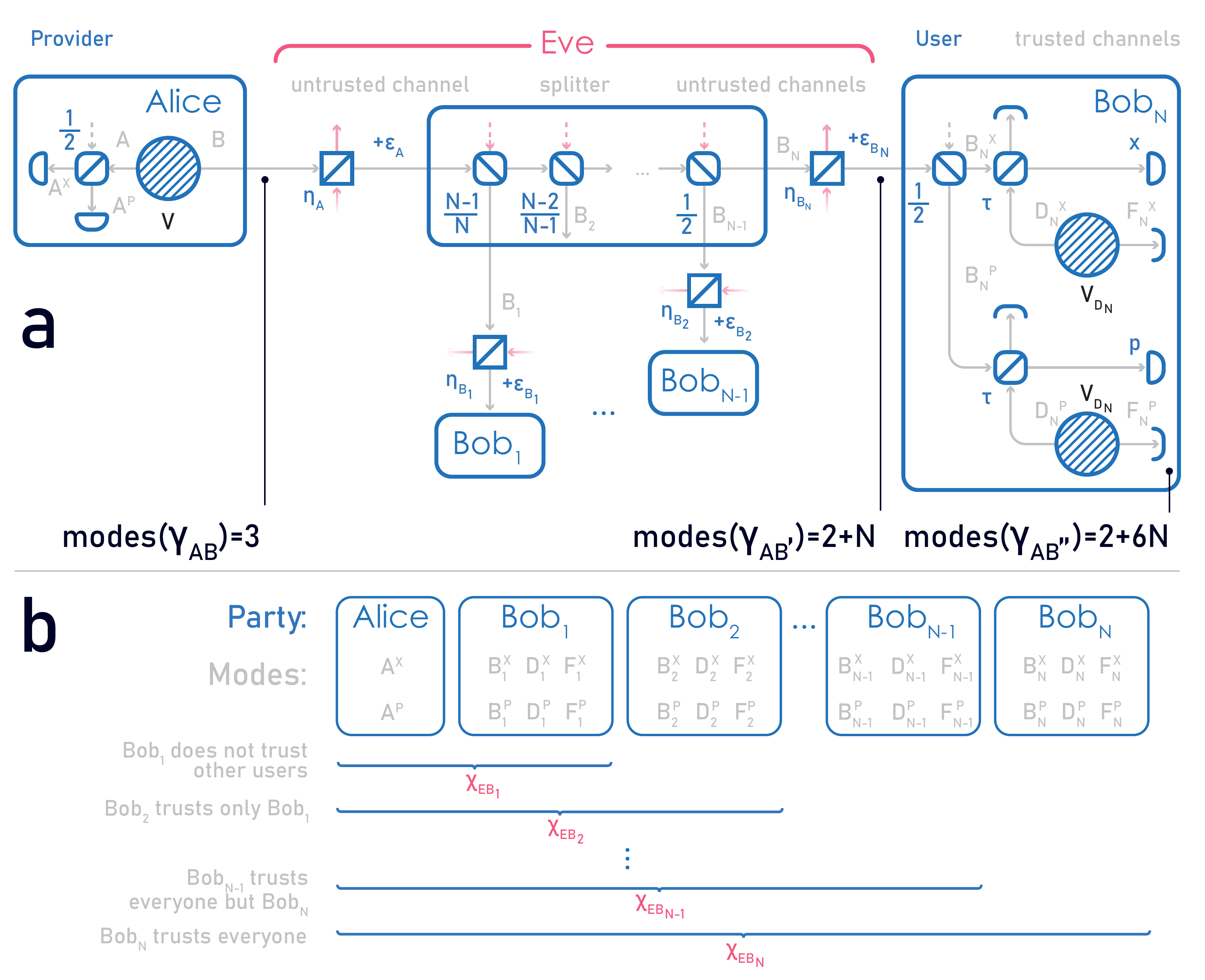}
    \caption{\textbf{CV-QPON protocol security analysis and trust assumptions}. (a) Broadcasting protocol entanglement-based scheme. (b) Successive user trust in trusted broadcasting protocol. }
    \label{fig:epr-based}
\end{figure}

The secure key rate for a Gaussian CV-QKD protocol can be evaluated using the covariance matrix of the equivalent entanglement-based setup due to the extremality of Gaussian states \citeS{wolf2006extremality}. Figure~\ref{fig:epr-based}(a) shows the  entanglement-based version of the prepare-and-measure CV-QPON protocol, where signal and noise sources are purified by a two-mode squeezed vacuum (TMSV) state described with a covariance matrix of the form
\begin{equation}
    \Gamma(V)=\begin{bmatrix}
V\mathbb{I} & \sqrt{V^2-1}\sigma_z  \\
\sqrt{V^2-1}\sigma_z & V\mathbb{I} 
\end{bmatrix}, 
\end{equation}
where $\mathbb{I}$ is a $2\times 2$ identity matrix, and Pauli matrix $\sigma_z=\text{diag}\left[1,0,0,-1\right]$. In this setup, Alice prepares a coherent state  for broadcasting by heterodyning one of the modes ($A$) of the source with covariance matrix $\Gamma(V)$. The initial signal state is given by the covariance matrix $\gamma_{AB}=(S_{1/2}\otimes\mathbb{I})\left[\mathbb{I}\otimes\Gamma(V)\right](S_{1/2}\otimes\mathbb{I})^T$, with a symplectic transformation $S_{1/2}$  corresponding to a balanced beamsplitter operation. The model of the source can be more involved, accommodating varying levels of modulation for each quadrature, limited squeezing and/or preparation noise\citeS{usenko2011squeezed, derkach2017continuous}. \par

The broadcasted signal (mode $B$) propagates through a quantum channel characterized by transmittance $\eta_A$ and excess noise variance  $\varepsilon_A$. The signal mode $B$ is then split into $N$ modes $B_1,\dots B_N$, for each respective user. These modes travel through quantum channels with distinct transmittances $\eta_{B_1}\dots\eta_{B_N}$ and noise variances $\varepsilon_{B_1}\dots\varepsilon_{B_N}$. Typically, the variance of excess noise might vary between the different quadratures. However, in our case, the variances in both quadratures were sufficiently similar, allowing us to simplify our analysis by using their averaged values. In general, Eve's  
ancillary modes can be correlated, enabling her to extract more information from the broadcasted states during individual measurements. However, we assume all quantum channels are independent, implying that the excess noise is also independent and added after the signal has been split for distribution to users (i.e., $\varepsilon_A=0$). This is because the influence of noise added before the splitter $\varepsilon_A$ would drastically decrease with a growing number of users $N$. The exploration of more sophisticated attack strategies, which might not adhere to these simplifications, is deferred to future research. 
After the signal passes through the quantum channels, the resulting covariance matrix, denoted as $\gamma_{AB'}$,  contains $2+N$ modes ($A^{x(p)},\,B_1\dots B_N$):
\begin{equation}
    \gamma_{AB'}= \frac{1}{2}\begin{bmatrix}
\gamma_{A^xA^p} & \sqrt{2\eta_1(V^2-1)}\varsigma_z & \cdots & -\sqrt{2\eta_N(V^2-1)}\varsigma_z \\
\sqrt{2\eta_N(V^2-1)}\varsigma_z^T & \{(V-1)\eta_1 +2 + \varepsilon_1 \}\mathbb{I} & & -\sqrt{2\eta_1\eta_N}(V-1)\mathbb{I} \\
\vdots & & \ddots & \vdots \\
-\sqrt{2\eta_N(V^2-1)}\varsigma_z^T & -\sqrt{2\eta_1\eta_N}(V-1)\mathbb{I} & \dots & \{(V-1)\eta_N +2 + \varepsilon_N \}\mathbb{I}
\end{bmatrix}, \text{with } \varsigma_z=\begin{bmatrix}
1 & 0 \\
0 & -1 \\
1 & 0 \\
0 & -1 
\end{bmatrix},  
\end{equation}
where $\eta_i=\eta_A\eta_{B_i}/N$ is the total transmittance, $\varepsilon_i=\varepsilon_{B_i}$, and $\gamma_{A^xA^p}=\frac{1}{2}\text{diag}\left[(V+1)\mathbb{I}, (V-1)\mathbb{I}, (V-1)\mathbb{I},(V+1)\mathbb{I},\right]$. \par

On the receiving end, each user splits the incoming signal (mode $B_i$) on a balanced beam splitter into two modes ($B_i^x$ and $B_i^p$) and measures them with imperfect homodyne detectors in respective quadratures. The imperfect detection is modeled as linear interaction on a beam splitter with transmitance $\tau$. This interaction is accompanied by a thermal noise of variance $\nu$, corresponding to electronic, and purified by a TMSV source (modes $D_i$ and $F_i$) described by a covariance matrix $\Gamma(V_{D_i})$, with $V_{D_i} = 1 + \nu_i/(1-\tau)$. The trusted detection noise can be correlated between quadratures; regardless, it will not contribute to the knowledge of Eve. In scenarios where heterodyne measurements are imbalanced or efficiencies vary by quadrature, the state of Bob$_i$ can be described by a 6 modes covariance matrix containing modes $B_i^{x(p)}$, $D_i^{x(p)}$, and $F_i^{x(p)}$. Here, variances of electronic noise $\nu_i$ were averaged over quadratures at each Bob$_i$, similarly as the excess noise $\varepsilon$, and detection efficiencies were common for all users $\tau=\tau_1=\dots=\tau_N$. Bob$_i$ homodynes the state with variance $V_{B_i^x}=1+\left[(V-1)\eta_i+\varepsilon_i\right]\frac{\tau}{2}+\nu_i$ in the mode $B_i^x$, and the covariance matrix of modes $B_i^x$, $D_i^x$ and $F_i^x$ after the interactions can be written as:

\begin{equation}
    \gamma_{B_i^xD_i^xF_i^x}=\begin{bmatrix}
V_{B_i^x}\mathbb{I} & \tau\frac{2\nu_i-(1-\tau)(\varepsilon_i+(V-1)\eta_{B_i})}{2\sqrt{\tau(1-\tau)}}\mathbb{I} & \sqrt{\frac{\nu_i^2 +2\nu_i\left(1- \tau\right)}{(1-\tau)}} \sigma_z \\
\tau\frac{2\nu_i-(1-\tau)(\varepsilon_i+(V-1)\eta_{B_i})}{2\sqrt{\tau(1-\tau)}}\mathbb{I} & \frac{1-\tau}{2}\left(2+\varepsilon_i+(V-1)\eta_{B_i}\right)+\tau V_{D_i}\mathbb{I} & \sqrt{\tau\nu_i\frac{\nu+2(1-\tau)}{(1-\tau)^2}}\sigma_z \\
\sqrt{\frac{\nu_i^2 +2\nu_i\left(1- \tau\right)}{(1-\tau)}} \sigma_z & \sqrt{\tau\nu_i\frac{\nu_i+2(1-\tau)}{(1-\tau)^2}}\sigma_z & V_{D_i}\mathbb{I} \\
\end{bmatrix}. 
\end{equation}
The mode $B_i^x$ is correlated to mode $B_i^p$ as 
$\left[\eta_i\varepsilon_i-\eta_{B_i}(V-1)(1-\eta_A)\right]\sqrt{\frac{\tau}{2(1-\eta_i)\eta_i}}$, and 
with $C^{x}_{i,j}=B_j^x$ as $-\frac{\tau}{2}(V-1)\sqrt{\eta_{i}\eta_{j}}$. With $N$ users the overall broadcasted state will be contained in up to $2+6N$ modes, with the covariance matrix $\gamma_{AB''}$ that will serve a foundation for the security analysis of network protocols. 

   \section{Security analysis}  
The secure key rate between Alice and any Bob$_i$ is determined by Devetak-Winters formula \citeS{devetak2005distillation}:
    \begin{equation} \label{eq:key0}
        K_i(\eta,\varepsilon)=\text{max}\left[0,\beta_i I_{AB_i} - \chi_{EB_i}\right],
    \end{equation}
where both the mutual information $I_{AB_i}$ and the Holevo bound $\chi_{EB_i}$ can be evaluated based on the covariance matrix $\gamma_{AB''}$. The former is the same for time-sharing and broadcasting protocols and is commonly determined by signal-to-noise ratio (SNR) as,
\begin{equation}
    I_{AB_i}=\frac{1}{2}\log_2\left[1+\frac{\frac{\eta_i\tau}{2}(V-1)}{1+\nu_i+\varepsilon_i\frac{\tau}{2}}\right].
\end{equation}
The mutual information between users $$I_{B_iB_j}=\frac{1}{2}\log_2\left[\frac{V_{B_i^x}}{V_{B_i^x}-\frac{(C^{x}_{i,j})^2}{V_{B_j^x}}}\right],$$ corresponding to passive state preparation \citeS{qi2018passive}, which quickly deteriorates with increasing number of users or extending distance from the provider.\par 
In asymptotic regime, where collective attacks have been shown to be optimal \citeS{leverrier2015composable, leverrier2017security}, Eves' knowledge on the measurement results of the reference user is upper bounded by the Holevo bound $\chi_{EB_i}=S(\rho_E)-S(\rho_{E|B_i})$ \citeS{holevo2001evaluating}. Eve holds the purification of the shared trusted state $\rho_{AB_iC}$, with $tr(\rho_{AB_iCE})=1$, where state $\rho_C$ corresponds to the part of broadcasted signal distributed to other non-reference users (and respective trusted measurements). Consequently, the bound is simplified to $\chi_{EB_i}=S(\rho_{AB_iC})-S(\rho_{AC|B_i})$,  and can be evaluated based on the re-constructed covariance matrix $\gamma_{AB''}$. \par
While the sub-systems of Alice $\rho_A$ and reference user $\rho_{B_i}$ (which also holds the purification of relevant trusted noise and loss) are settled, the main difference between the protocols is the degree of trust that is defined by what part of the signal is assigned to sub-system $\rho_C$ (or equivalently to $\rho_E$). For time-sharing and untrusted broadcast protocols, all users are treated as part of Eve's system, i.e., all parts of the signal not received by the reference user are assumed to be part of $\rho_E$. In this case, the state $\rho_{AB_iC}$ is described by the covariance matrix containing modes $A_{x(p)}$, $B_i^{x(p)}$, $D_i^{x(p)}$, and $F_i^{x(p)}$. On the other hand, when the particular user (Bob$_j$) is regarded as trusted, modes $B_j^{x(p)}$, $D_j^{x(p)}$ and $F_j^{x(p)}$ are now assigned to the $\rho_{C}$ sub-system, leading to a decrease in Eve's knowledge. When all users are assumed to be trusted then a full $\gamma_{AB''}$ covariance matrix with $2+6N$ modes is used for the evaluation of the Holevo bound. A user operation is identical to the one in PTP protocols, and the reconstruction of the total covariance matrix along with determining individual Holevo bounds is up to the Provider.\par

    \subsection{Comparison of protocols}
In both, the time-sharing and untrusted broadcasting protocols, the user's key rate will be the same, potentially achieving the maximum limit set by the PLOB bound \citeS{PLOB}, which is determined by the channel transmittance $\eta_i$. The main distinction between these protocols lies in their ability to generate keys simultaneously. While the raw keys shared among users are not independent as $I_{B_iB_j}>0$, privacy amplification processes effectively decouple the keys from Eve, whose state also presumably contains all other users. Consequently,  final keys are completely independent among users. \par

\begin{figure}
    \centering
    \includegraphics[width=0.75\linewidth]{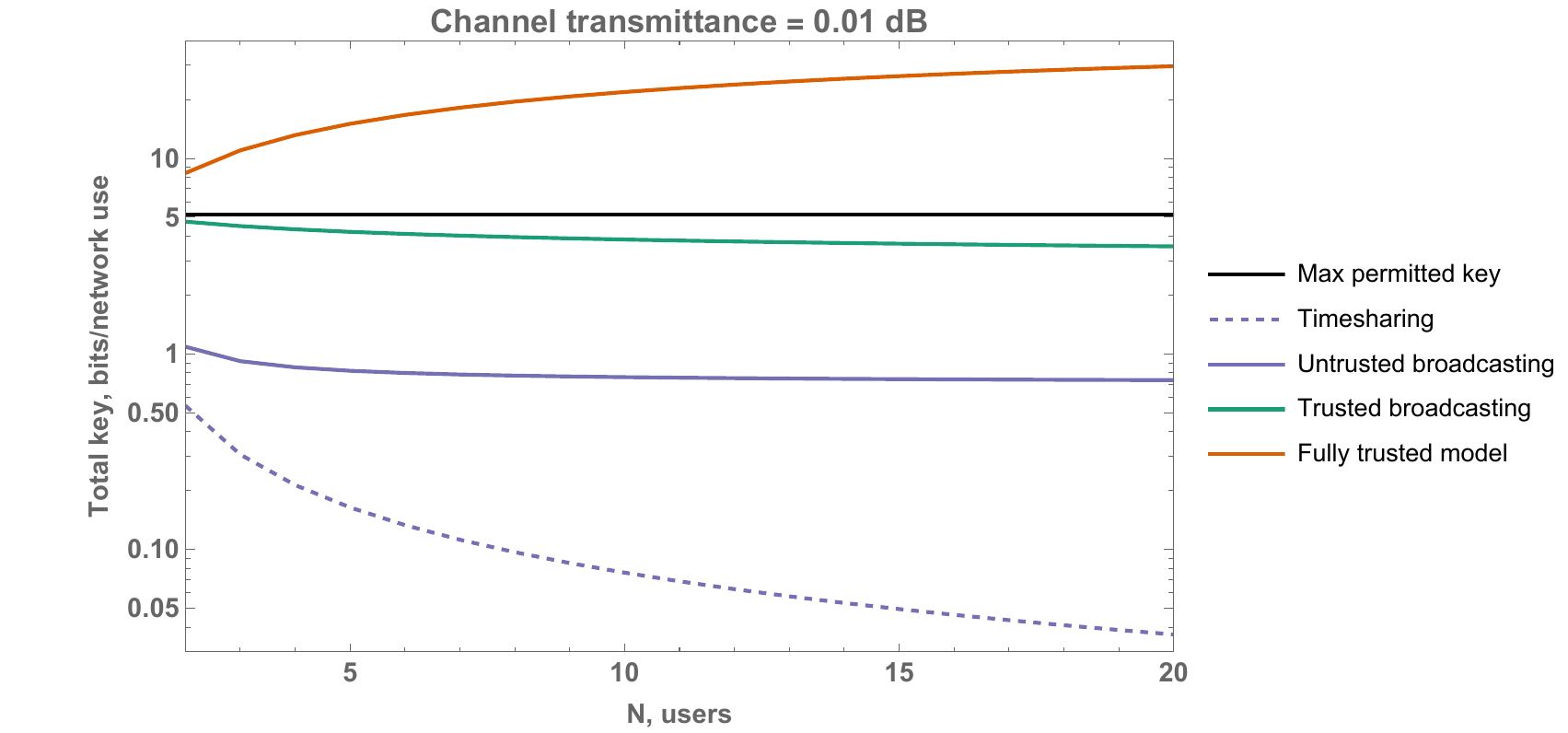}
    \caption{\textbf{Total network key comparison between protocols}. Noiseless channels $\varepsilon=0$, squeezing  modulation variance $V_{x(p)}=100$ SNU, perfect reconciliation $\beta=100\%$ and detectors $\tau=100\%$, $\nu=0$. }
    \label{fig:comparison}
\end{figure}

The trusted broadcasting protocol enhances the individual key rate of each user by relying on the faithful operation of network users. As illustrated in Fig.~\ref{fig:epr-based}(b), each user is allowed to designate a unique subset of other users as trusted. The differences between  protocols become apparent in conditions of extreme proximity to the Provider. In Fig.~\ref{fig:comparison} all protocols are compared to a standard point-to-point protocol with a single receiver (i.e. without the splitter $N=1$) and a channel transmittance of $\eta_A\eta_{B_1}=0.01$ dB—which sets the benchmark for the highest possible key rate. In scenarios where the number of users $N$ increases, the time-sharing approach experiences a rapid decline in key rate, whereas broadcasting protocols demonstrate resilience in maintaining a level of total key generation regardless of the network size. The signal loss significantly impacts the untrusted protocol, due to considering additional users as part of Eve's state. Conversely, trusting all users can result in underestimating Eve's potential knowledge $\chi_{EB_i}$, given the high correlation among all raw keys; knowledge of one key can offer insights into the others. This suggests that a comprehensive trust approach necessitates additional costs for privacy amplification. By allowing the number of trusted users to vary for each key, the trusted broadcasting protocol effectively circumvents this problem. \par
\begin{figure}
    \centering
    \includegraphics[width=0.85\linewidth]{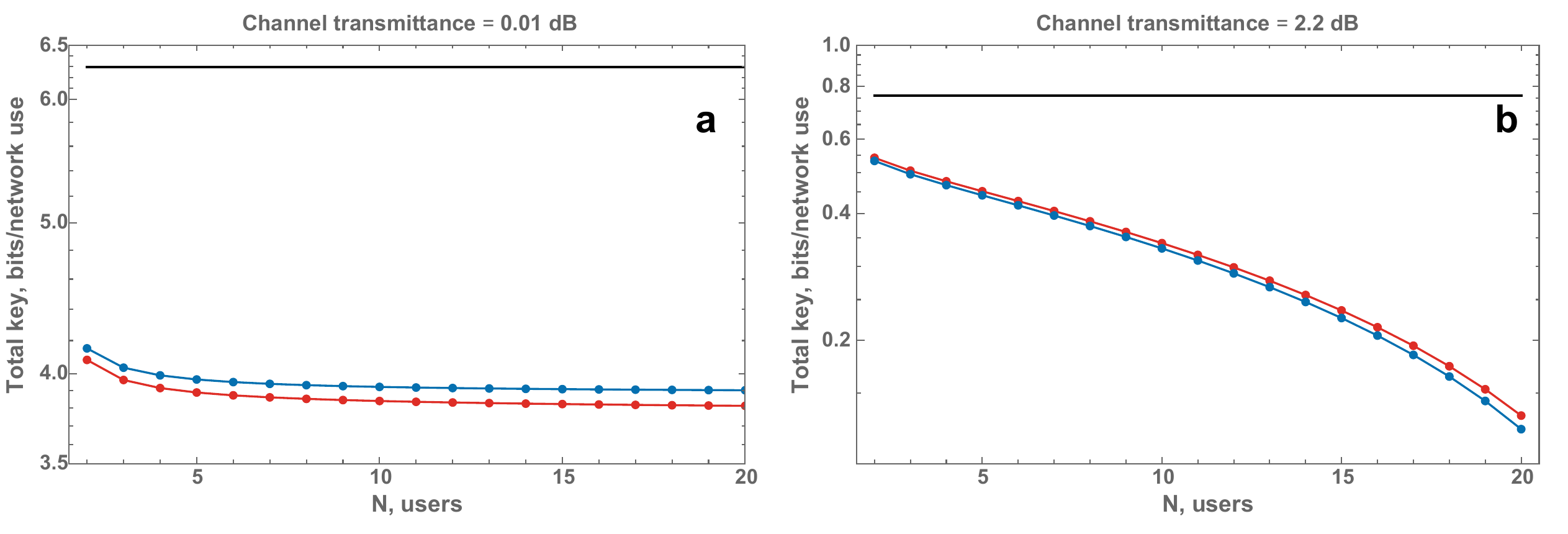}
    \caption{\textbf{Total network key comparison between squeezed-state trusted broadcasting (red) and state-merging-based (blue) protocols with maximal permitted key (black)}. (a) Noiseless channels $\varepsilon=0$,  modulation variance $V_{x(p)}=100$ SNU, perfect reconciliation $\beta=100\%$ and detectors $\tau=100\%$, $\nu=0$. (b) Channel noise $\varepsilon=0.5\%$ SNU, modulation variance $V_{x(p)}=4$ SNU, reconciliation efficiency $\beta=95\%$, detector efficiency $\tau=86\%$, electronic noise $\nu=2\%$ SNU. }
    \label{fig:squeezed}
\end{figure}

Extending the idea to the squeezed-state protocol, we note that the performance of the trusted broadcasting protocol is comparable with the broadcast protocol based on quantum state merging \citeS{takeoka2017}. The latter regards all users as untrusted when estimating the Holevo bound, but successively increases the joint mutual information between users. Both protocols provide a similar key rate, as shown in Fig.~\ref{fig:squeezed}, with a minor advantage of the state-merging-based protocol in the absence of trusted detection noise. Such noise effectively decouples Eve not only from a particular user but also from all other users as well. However, this cannot be taken into account during successive state merging as electronic noise has the same effect on mutual information regardless of it being trusted or not.  

  \subsection{Information reconciliation}

The information reconciliation (IR) process utilized a multi-dimensional (MD) reconciliation scheme, employing a multi-edge-type low-density-parity-check (MET-LDPC) code with a rate of 0.01 and codeword length $n= 8.192\times10^5$~\citeS{mani2021multiedge}. This code is theoretically designed to operate at a SNR of 0.007. However, in the context of  CV-QPON, users experience varying channel transmittances, which results in differing received SNRs. Consequently, the efficiency of the IR  exhibits significant variation across different users. \par
To enhance the performance of IR for all users within CV-QPON, we adopted a rate-adaptive reconciliation protocol. This approach allows for the flexible adjustment of the MET-LDPC code's rate according to the specific received SNR of each user. Through the application of puncturing and shortening techniques~\cite{wang2017efficient}, we can effectively modify the code rate either by increasing or decreasing it. The adjusted code rate is defined as, $R_{punc}=k/(n-p)$, for puncturing and $R_{sh}=(k-s)/(n-s)$ for shortening, where $k$ represents the number of information bites and  $p$ and $s$ are the lengths of puncturing and shortening, respectively. Table~\ref{tab:results-2} shows the modified rates along with the corresponding efficiency and frame error rate (FER) for each user. Most users deployed puncturing as their received SNR was above 0.007. On the other hand, despite operating at an SNR of approximately 0.007, Bob$_{6}$ had to employ a shortening strategy. This is because, for the code with finite length, the code rate needs to be strictly smaller than the channel capacity. \par  The selection of puncturing and shortening lengths cannot be made arbitrarily due to the inherent trade-off between FER performance and efficiency $\beta$~\citeS{hajomer2022modulation,hajomer2024long,jain2022practical}. Increasing the code rate through puncturing—by removing information—elevates the FER because the likelihood of incorrect frame decoding rises~\cite{martinez2012blind}. Conversely, shortening the code has the opposite effect. To implement these techniques, we developed a framework capable of processing 5.9 million symbols per second. This level of performance was facilitated by utilizing an NVIDIA GeForce RTX 2060 Mobile graphics processing unit (GPU) with the system consuming 2.5 gigabytes of memory. 

\begin{table}[t]
\caption{\textbf{Performance of adaptive reconciliation protocol}. The original MET-LDPC code has a rate $R = 0.01$, with codework length $n= 8.192\times10^5$ and information bits $k = 8192$.}
    \centering
    \resizebox{0.5\hsize}{!}{
    \begin{tabular}{|l|l|l|l|l|l|l|l|}
\hline
User          &SNR  &$p$ &$s$& $R_{pu}$&  $R_{sh}$& $\beta$,\%& FER,\%\\ \hline
\text{Bob}$_1$&0.0077& 10000& --& 0.0101& --&90.79 & 4.5 \\ \hline
\text{Bob}$_2$&0.0088& 130000& --&0.0118& --&93.23 & 43  \\ \hline
\text{Bob}$_3$&0.0091& 140000& --&0.0120 & --&91.37 & 22.3 \\ \hline
\text{Bob}$_4$&0.0083& 70000& --&0.0109& --&91.5  & 15.3 \\ \hline
\text{Bob}$_5$&0.0096&170000& --&0.0126& --&91.44 & 13.6 \\ \hline
\text{Bob}$_6$&0.00708&--& 550 & --  & 0.0093&91.9 & 21.5 \\ \hline
\text{Bob}$_7$&0.0082& 90000& --&0.0112& --&94.8  & 55.4 \\ \hline
\text{Bob}$_8$ &0.0097& 170000& --&0.0126& --&90.78 & 9.5 \\ \hline
\end{tabular}
}
\label{tab:results-2}
\end{table}

 \subsection{Correlation of users}
\begin{figure}[t]
\centering
\includegraphics[width=.5\linewidth]{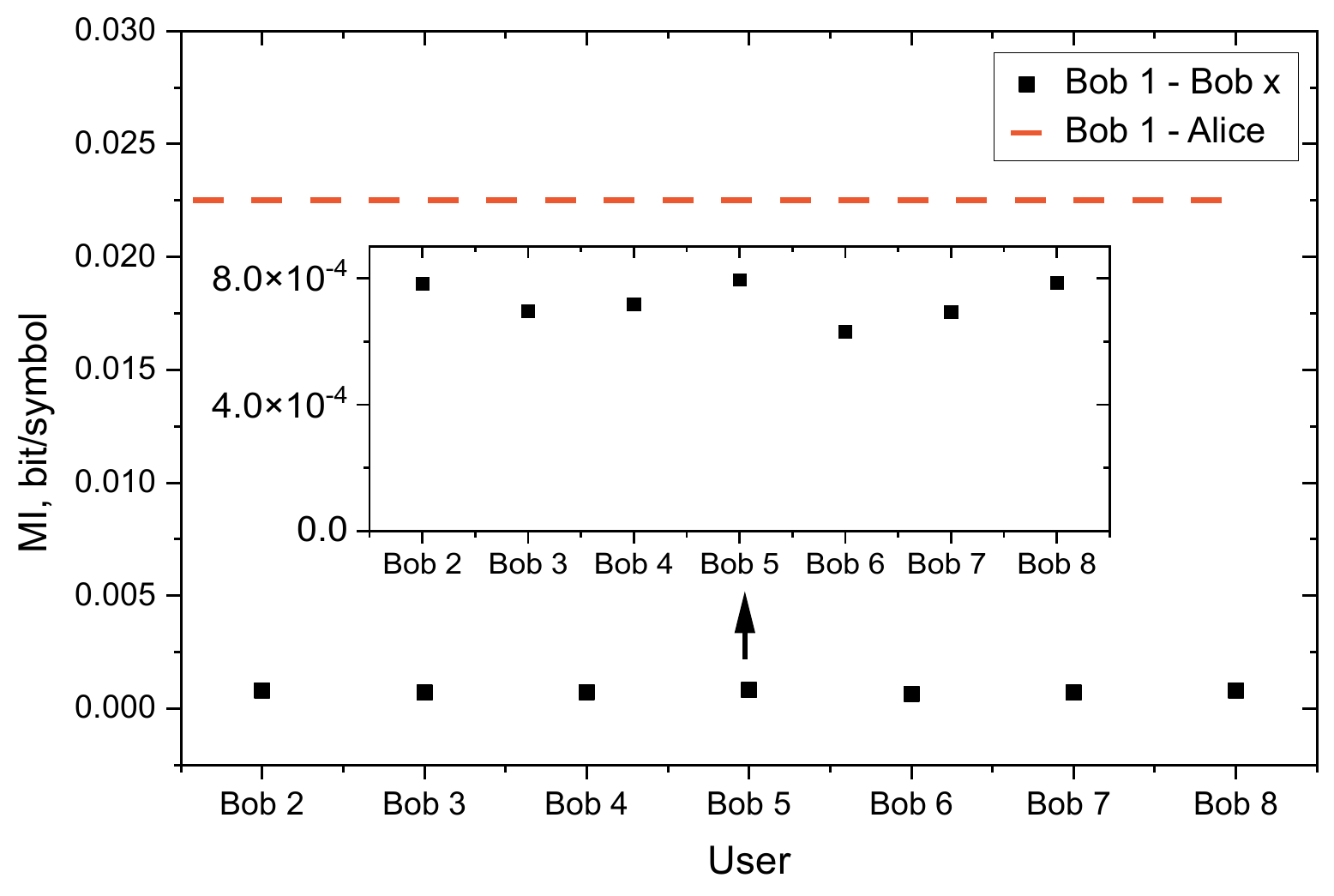}
\caption{\textbf{Analysis of user correlations within the network}. Experimentally obtained mutual information (MI) between Bob$_1$ and other network users compared to the mutual information between Bob$_1$ and the provider, Alice. The inset shows the MI with a different scale.}  
\label{fig:4}
\end{figure}

To demonstrate our CV-QPON protocols' ability to enable simultaneous key generation among users, we examined correlations of the users' measurement results. Figure~\ref{fig:4}  shows the mutual information (MI) between $\text{Bob}_1$ and other users, revealing a low correlation, as indicated by small MI values (highlighted by a black square in the figure inset). The rather small correlations are largely due to the independently measured quantum noise by each user.  In contrast, the MI between Alice and $\text{Bob}_1$ is substantially higher, by two orders of magnitude compared to other users, as represented by the red dashed line. Therefore, implementing reverse reconciliation~\cite{grosshans2003quantum} allows  Alice and each user to gain an information advantage, thus facilitating the concurrent reconciliation of independent keys, which is a distinctive aspect of the CV-QPON protocols.

  \begin{figure}
    \centering
    \includegraphics[width=0.5\linewidth]{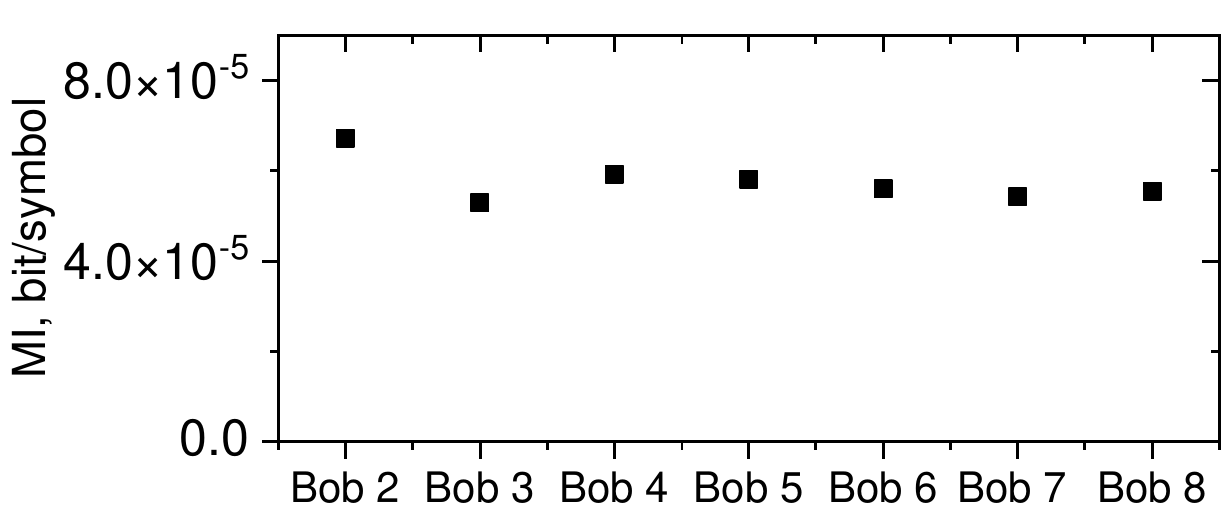}
    \caption{\textbf{Analysis of noise correlation measured by different users.} The excess noise of $\text{Bob}_1$ is considered as a reference for mutual information measurement.}
    \label{fig:noise correlation}
\end{figure}

To conduct the analysis of correlations between noise measured by different users, we infer the total noise variables based on Bob's measurement as,  
\begin{equation}
     \xi^{x(p)}_{l} = B_{l}^{x(p)} - g~\times A^{x(p)}.
\end{equation}
Here, $g$ represents a scaling factor, which can be calculated as $\text{Cov}(B_{l}^{x(p)} ,A^{x(p)})/\text{Var}(A^{x(p)})$, where $\text{Cov}$ and $\text{Var}$ are covariance and variance, respectively. Figure~\ref{fig:noise correlation} displays the mutual information between the noise measured by Bob$_1$ and another user. The mutual information between the noise measurements across users is an order of magnitude lower than observed among users. This residual correlation can be attributed to the finite-size effects of these Gaussian variables. Thereby, the noise measured by users is largely independent, and any observed correlation among users can mainly be attributed to Alice's modulation, as depicted in Fig~\ref{fig:4}. These results support our assumption of noise independence, aligning closely with practical implementations.

 \newpage
\bibliographystyleS{unsrt}
\bibliographyS{libsup}

\end{document}